\pdfoutput=1
\documentclass{PoS}

\usepackage{amsfonts,amsmath,graphicx,bm,slashed,color,mathrsfs,placeins}

\definecolor{darkgreen}{rgb}{0.0,0.3,0.0}

\newcommand{\bss}[1]{\ensuremath{{\boldsymbol{#1}}}}
\newcommand{\red}[1]{{\textcolor{red}{#1}}}
\newcommand{\green}[1]{{\textcolor{darkgreen}{#1}}}

\title{Lattice QCD calculation of form factors for $\Lambda_b \to \Lambda(1520) \ell^+ \ell^-$ decays}

\ShortTitle{Lattice QCD calculation of form factors for $\Lambda_b \to \Lambda(1520) \ell^+ \ell^-$ decays}

\author{\speaker{Stefan Meinel}\\
        Department of Physics, University of Arizona, Tucson, AZ 85721, USA\\
        RIKEN BNL Research Center, Brookhaven National Laboratory, Upton, NY 11973, USA\\
        E-mail: \email{smeinel@email.arizona.edu}}

\author{Gumaro Rendon\\
        Department of Physics, University of Arizona, Tucson, AZ 85721, USA\\
        E-mail: \email{jgrs@email.arizona.edu}}

\abstract{Experimental results for mesonic $b \to s \mu^+ \mu^-$ decays show a pattern of deviations
from Standard-Model predictions, which could be due to new fundamental physics or due to
an insufficient understanding of hadronic effects. Additional information on the $b \to s \mu^+ \mu^-$
transition can be obtained from $\Lambda_b$ decays. This was recently done using the process
$\Lambda_b \to \Lambda \mu^+ \mu^-$, where the $\Lambda$ is the lightest strange baryon. A further
interesting channel is $\Lambda_b \to p^+ K^- \mu^+ \mu^-$,
where the $p^+ K^-$ final state receives contributions from multiple higher-mass $\Lambda$ resonances.
The narrowest and most prominent of these is the $\Lambda(1520)$, which has $J^P=\frac32^-$.
Here we present an ongoing lattice QCD calculation of the relevant $\Lambda_b \to \Lambda(1520)$ form factors.
We discuss the choice of interpolating field for the $\Lambda(1520)$, and explain our method for extracting
the fourteen $\Lambda_b \to \Lambda(1520)$ helicity form factors from correlation functions
that are computed in the $\Lambda(1520)$ rest frame. We present preliminary numerical results
at a pion mass of 340 MeV and a lattice spacing of $0.11$ fm. This calculation uses a domain-wall
action for the $u$, $d$, and $s$ quarks and a relativistic heavy-quark action for the $b$ quark,
and is based on gauge-field configurations generated by the RBC and UKQCD Collaborations.
}

\FullConference{34th annual International Symposium on Lattice Field Theory\\
		24-30 July 2016\\
		University of Southampton, UK}

\begin{document}

\FloatBarrier
\section{Introduction}
\FloatBarrier

\label{sec:introduction}

Flavor-changing neutral-current decays of bottom hadrons play an important role in the search for physics beyond the Standard Model.
The effective Hamiltonian describing $b \to s \ell^+\ell^-$ decays at low energies \cite{Grinstein:1988me} contains the operators
\begin{equation}
    O_{7(7')}  = \frac{m_b}{e}\bar{s} \sigma^{\mu\nu} P_{R(L)} b \: F_{\mu\nu}, \hspace{1ex}  O_{9(9')}  = \bar{s} \gamma_\mu P_{L(R)} b\: \bar{\ell} \gamma^\mu \ell, \hspace{2ex}    O_{10(10')}  = \bar{s} \gamma_\mu P_{L(R)} b\:\bar{\ell} \gamma^\mu \gamma_5 \ell,
\end{equation}
as well as four-quark and gluonic operators. The Wilson coefficients $C_i$ of these operators encode the short-distance physics and can be computed perturbatively
in the Standard Model and in various new-physics scenarios. The values of $C_i$ can also be constrained by fitting the decay rates and angular distributions measured in experiments,
provided that the relevant hadronic matrix elements are known. Global analyses of experimental data for mesonic $b \to s \mu^+\mu^-$ decays, which use a combination of several theoretical methods
[including lattice QCD for $O_{7(7')}$, $O_{9(9')}$, and $O_{10(10')}$], yield best-fit values for $C_9$ that are approximately 25\% below the Standard-Model prediction
(see, e.g., Refs.~\cite{Altmannshofer:2014rta,Altmannshofer:2015sma,Descotes-Genon:2015uva,Hurth:2016fbr}). However, the results for $C_9$ also depend
on nonlocal matrix elements involving the four-quark operators $O_1$ and $O_2$, which are enhanced by charmonium resonances, and the approximations
used for these matrix elements need further scrutiny.

\begin{table}[b]
\begin{tabular}{|lcccc|}
 \hline
 & { \hspace{-2ex} Probes all \hspace{-2ex}        }   & {  Final hadron  }  & { \hspace{-2ex} Charged hadrons from \hspace{-2ex}  } & { LQCD } \\
 & { \hspace{-2ex} Dirac structures \hspace{-2ex}  }   & {  QCD-stable    }  & {  $b$-decay vertex     } & { Refs.} \\
\hline
 { $B^+ \to K^+ \ell^+ \ell^-$ \hspace{-4ex} }                                   & \red{$\times$}          & \green{$\checkmark$}    & \green{$\checkmark$}  &  { \cite{Bouchard:2013pna,Bouchard:2013mia,Bailey:2015dka,Du:2015tda}}  \\
 { $B^0 \to K^{*0}(\to K^+\pi^-) \ell^+ \ell^-$ \hspace{-4ex} }                  & \green{$\checkmark$}    & \red{$\times$}          & \green{$\checkmark$}  &  { \cite{Horgan:2013hoa,Horgan:2013pva,Flynn:2015ynk}} \\
 { $B_s \to \phi(\to K^+K^-) \ell^+ \ell^-$ \hspace{-4ex} }                      & \green{$\checkmark$}    & \red{$\times$}          & \green{$\checkmark$}  &  { \cite{Horgan:2013hoa,Horgan:2013pva,Flynn:2015ynk}} \\
 { $\Lambda_b^0 \to \Lambda^0 (\to p^+\pi^-)\, \ell^+ \ell^-$ \hspace{-4ex}}     & \green{$\checkmark$}    & \green{$\checkmark$}    & \red{$\times$}        &  { \cite{Detmold:2012vy,Detmold:2016pkz,Meinel:2016grj}} \\
 { $\Lambda_b^0 \to \Lambda^{*0} (\to p^+ K^-)\, \ell^+ \ell^-$ \hspace{-4ex}}   & \green{$\checkmark$}    & \red{$\times$}          & \green{$\checkmark$}  &  { This work}  \\
\hline
\end{tabular}
\caption{\label{tab:decaychannels}Comparison of exclusive $b\to s\ell^+\ell^-$ decay channels.}
\end{table}

In addition to the commonly studied $B$ and $B_s$ decays, the $b \to s \ell^+\ell^-$ couplings can also be probed in decays of $\Lambda_b$ baryons
(see Table \ref{tab:decaychannels} for a comparison of the most important semileptonic decay modes). Recently, the authors of Ref.~\cite{Meinel:2016grj}
included, for the first time, the LHCb results for the differential branching fraction and three angular observables
of the decay $\Lambda_b \to \Lambda (\to p^+\pi^-)\, \mu^+ \mu^-$ \cite{Aaij:2015xza} in an analysis of the Wilson coefficients $C_{9,9',10,10'}$. From
a theoretical point of view \cite{Detmold:2016pkz, Boer:2014kda}, this decay combines the best aspects of $B \to K \ell^+ \ell^-$ (having only a single
QCD-stable hadron in the final state, which simplifies the lattice QCD calculation of the form factors) and $B \to K^*(\to K\pi) \ell^+\ell^-$
(providing a large number of observables that give full sensitivity to all Dirac structures in the effective Hamiltonian). The fits performed
in Ref.~\cite{Meinel:2016grj} prefer a positive shift in $C_9$, contrary to previous fits of only mesonic decays. This behavior could hint at large duality
violations in the high-$q^2$ operator product expansion that is used to approximate the nonlocal matrix elements of $O_1$ and $O_2$. Unfortunately, the statistical
uncertainties in the $\Lambda_b \to \Lambda (\to p^+\pi^-)\, \mu^+ \mu^-$ data \cite{Aaij:2015xza} are still quite large. One experimental challenge with this decay is that the hadron in
the final state, the lightest $\Lambda$ baryon, is electrically neutral and long-lived. It is therefore worth exploring decays proceeding through unstable
$\Lambda^*$ resonances, which can immediately decay into charged particles such as $p^+ K^-$ and produce tracks in the particle detectors that originate
from the $b$-decay vertex.

The $p^+ K^-$-invariant-mass distribution in $\Lambda_b \to p^+ K^-\, \mu^+ \mu^-$ decays is expected to be similar to that in $\Lambda_b \to p^+ K^-\:J/\psi$.
As can be seen in Fig.~3 of Ref.~\cite{Aaij:2015tga}, a large number of $\Lambda^*$ resonances contribute to this decay in overlapping mass regions. However, one
resonance produces a narrow peak that clearly stands out above the other contributions: the $\Lambda(1520)$, which is the lightest resonance with $J^P=\frac32^-$.
The $\Lambda(1520)$ has a width of $15.6\pm 1.0$ MeV \cite{Agashe:2014kda} and appears in the coupled channels $pK$, $\Sigma\pi$, $\Lambda\pi\pi$, and, less importantly, $\Sigma\pi\pi$.
Given the small width, a naive analysis in which the $\Lambda(1520)$ is treated as if it were QCD-stable is expected to be quite accurate, and is therefore
justified in a first lattice QCD calculation of $\Lambda_b \to \Lambda(1520)$ form factors. When working in the $\Lambda(1520)$ rest frame, the lowest energy level in the finite
lattice volume can be identified with the resonance in the narrow-width approximation; in the rest frame, the $pK$, $\Sigma\pi$, $\Lambda\pi\pi$, and $\Sigma\pi\pi$ scattering-like states
will appear at higher energies due to the nonzero back-to-back momenta required for a coupling to $J^P=\frac32^-$.

In the following, we will use the notation $\Lambda^*$ to refer to the $\Lambda(1520)$. The $\Lambda_b\to\Lambda^*$ matrix elements of the $b\to s$ vector, axial vector, and tensor currents [as
needed for $O_{7(7')}$, $O_{9(9')}$, and $O_{10(10')}$] are described by 14 form factors \cite{Mott:2011cx}. Following the approach of Ref.~\cite{Feldmann:2011xf}, we have
derived a new helicity-based definition of the $\Lambda_b\to\Lambda^*$ form factors. The decomposition for the vector current reads

\vspace{2ex}
$ \displaystyle  \langle \Lambda^*(p^\prime,s^\prime) |\, \bar{s}\gamma^\mu b \, |  \Lambda_b(p,s) \rangle  $
\begin{eqnarray}
\nonumber &=&
\bar{u}_\lambda(p^\prime,s^\prime) \Bigg[ {f_0}\,\frac{(m_{\Lambda_b}-m_{\Lambda^*})\,p^\lambda q^\mu}{m_{\Lambda_b}\, q^2}   +\: {f_+}\,\frac{(m_{\Lambda_b}+m_{\Lambda^*})\, p^\lambda ( q^2(p^\mu+p^{\prime \mu}) - (m_{\Lambda_b}^2-m_{\Lambda^*}^2) q^\mu   )}{m_{\Lambda_b}\, q^2\, s_+}   \\
\nonumber && \hspace{8ex} +\: {f_\perp} \Bigg(\frac{p^\lambda \gamma^\mu}{m_{\Lambda_b}} - \frac{2\, p^\lambda(m_{\Lambda_b}p^{\prime \mu} + m_{\Lambda^*} p^\mu)}{m_{\Lambda_b}\,s_+}    \Bigg)   \\
 && \hspace{8ex} +\: {f_{\perp^\prime}} \Bigg( \frac{p^\lambda \gamma^\mu}{m_{\Lambda_b}} - \frac{2\, p^\lambda p^{\prime \mu}}{m_{\Lambda_b}m_{\Lambda^*}}
 + \frac{2\, p^\lambda(m_{\Lambda_b}p^{\prime \mu} + m_{\Lambda^*} p^\mu  )}{m_{\Lambda_b}\,s_+} + \frac{s_-\,  g^{\lambda\mu}}{m_{\Lambda_b}m_{\Lambda^*}}\Bigg) \Bigg] u(p, s), \hspace{3ex} \null
\end{eqnarray}
where $q=p-p^\prime$, $s_\pm=(m_{\Lambda_b}\pm m_{\Lambda^*})^2-q^2$, and the form factors $f_0$, $f_+$, $f_\perp$, $f_{\perp^\prime}$ are functions of $q^2$.
Above, $\bar{u}_\lambda(p^\prime,s^\prime)$ is the Rarita-Schwinger spinor for the $\Lambda^*$. Similar relations
are obtained for the currents $\bar{s}\,\gamma^\mu\gamma_5\, b$ (form factors $g_0$, $g_+$, $g_\perp$, $g_{\perp^\prime}$), $\bar{s}\,i\sigma^{\mu\nu}q_\nu\,b$ (form factors
$h_+$, $h_\perp$, $h_{\perp^\prime}$), and $\bar{s}\,i\sigma^{\mu\nu}\gamma_5q_\nu\,b$ (form factors $\widetilde{h}_+$, $\widetilde{h}_\perp$, $\widetilde{h}_{\perp^\prime}$).

\FloatBarrier
\section{Interpolating field for the $\Lambda(1520)$}
\FloatBarrier

\begin{table}[t]

\vspace{-2ex}

\begin{center}

\begin{tabular}{|ccllllc|}
\hline            &  &                                &                             &                                &                            &          \\[-2.5ex]
$N_s^3\times N_t$ & $\beta$  & $a\,m_{u,d}^{(\mathrm{sea})}$  & $a\,m_{s}^{(\mathrm{sea})}$ & $a\,m_{u,d}^{(\mathrm{val})}$  & $a\,m_s^{(\mathrm{val})}$  & $a$ [fm] \\
\hline
$24^3\times64^{\phantom{X^X}}$  \hspace{-3ex}   &  2.13  &0.005   & 0.04  & 0.005  & 0.0323    &  0.1106(3)   \\
\hline
\end{tabular}

\vspace{1ex}

\begin{tabular}{|ccccc|}
\hline
$m_\pi$ [MeV]  & $m_K$ [MeV]  & $m_N$ [MeV] & $m_\Lambda$ [MeV] & $m_\Sigma$ [MeV]  \\
\hline
340(1)     & 550(2)       & 1168(5)     & 1272(5)          & 1320(6)           \\
\hline
\end{tabular}

\caption{\label{tab:params}Lattice parameters and preliminary results for selected hadron masses. Details on the lattice actions and ensemble generation can be found in Ref.~\cite{Aoki:2010dy}.
We use all-mode-averaging (AMA) \cite{Shintani:2014vja} with 1 exact and 32 sloppy measurements per configuration.}
\end{center}
\end{table}

\begin{figure}[t]
\begin{center}
 \includegraphics[width=0.6\linewidth]{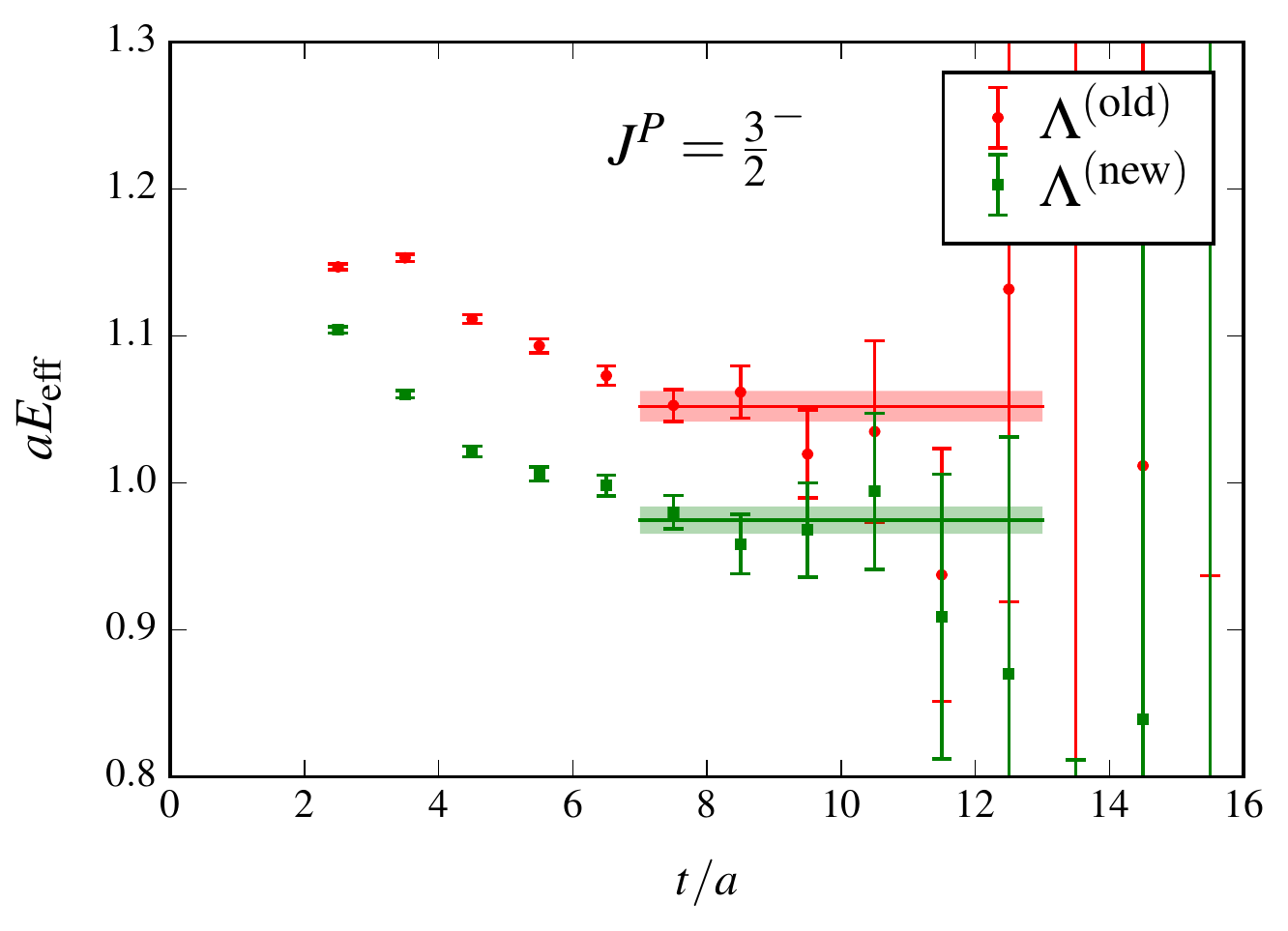}
 \caption{\label{fig:lambda2pt}Effective-energy plot for the two-point functions computed with the interpolating fields defined in Eqs.~(\protect\ref{eq:Lambda1}) and (\protect\ref{eq:Lambda2}). The preliminary results shown
 here are from 311 configurations (with AMA). The energies obtained from the fits are $1878(19)$ MeV and $1740(17)$ MeV.}
\end{center}
\end{figure}

We work in the $\Lambda(1520)$ rest frame to allow an exact projection to the $J^P=\frac32^-$ quantum numbers, and also for the reasons discussed in Sec.~\ref{sec:introduction}.
In a first (unsuccessful) attempt at calculating the form factors, we used the interpolating field
\begin{equation}
 \Lambda_{j\gamma}^{({\rm old})} = \epsilon^{abc}\:(C\gamma_j)_{\alpha\beta}\:\left( \tilde{u}^a_\alpha\:\tilde{s}^b_\beta\: \tilde{d}^c_\gamma - \tilde{d}^a_\alpha\:\tilde{s}^b_\beta\: \tilde{u}^c_\gamma \right), \label{eq:Lambda1}
\end{equation}
which has isospin 0 as required, and which we projected to $J^P=\frac32^-$ by contracting with $P^{kj} = \left(g^{kj}-\frac13\gamma^k\gamma^j\right) \frac{1-\gamma_0}{2}$ (above, the tilde on the quark fields denotes
gauge-covariant Gaussian \mbox{smearing}). With the interpolating field $\Lambda_{j\gamma}^{({\rm old})}$, the numerical results for the ratios of three-point and two-point functions used to extract the form factors were
very noisy and did not show plateaus. We then noticed that a previous lattice QCD study of $\Lambda$-baryon spectroscopy using interpolating fields
similar to Eq.~(\ref{eq:Lambda1}) in fact did not find a $\Lambda(1520)$-like state \cite{Engel:2012qp}, while the calculation of Ref.~\cite{Edwards:2012fx}, which included interpolating
fields with covariant derivatives, did. This can be understood from quark models, in which the $\Lambda(1520)$ dominantly has an $L=1$, $S=1/2$, and flavor-$SU(3)$ singlet structure \cite{Gromes:1982ze}, very different
from Eq.~(\ref{eq:Lambda1}). We therefore now use the interpolating field
\begin{equation}
\Lambda_{j\gamma}^{({\rm new})} = \epsilon^{abc}\:(C\gamma_5)_{\alpha\beta}\:\left[ \tilde{s}^a_\alpha\:\tilde{d}^b_\beta\: (\nabla_j \tilde{u})^c_\gamma  -  \tilde{s}^a_\alpha\:\tilde{u}^b_\beta\: (\nabla_j \tilde{d})^c_\gamma  + \tilde{u}^a_\alpha\:(\nabla_j \tilde{d})^b_\beta\: \tilde{s}^c_\gamma - \tilde{d}^a_\alpha\:(\nabla_j \tilde{u})^b_\beta\: \tilde{s}^c_\gamma  \right], \label{eq:Lambda2}
\end{equation}
which matches the structure suggested by  nonrelativistic quark models and has naturally negative parity, so that it can be projected to $J^P=\frac32^-$ by contracting with $P^{kj} = \left(g^{kj}-\frac13\gamma^k\gamma^j\right) \frac{1+\gamma_0}{2}$ (note the plus sign).
In Eq.~(\ref{eq:Lambda2}), covariant derivatives acting on the strange quark have been eliminated using ``integration by parts'' (which is possible only at zero momentum). Numerical
results for the two-point functions, with the lattice parameters given in Table \ref{tab:params}, are shown in Fig.~\ref{fig:lambda2pt}. The two-point function of the new interpolating field (\ref{eq:Lambda2})
shows a plateau at an energy close to $m_N+m_K$ and $m_\Sigma+m_\pi$, as expected for the $\Lambda(1520)$, while the two-point function of the old interpolating field (\ref{eq:Lambda1}) shows an apparent plateau at a
significantly higher energy that is likely associated with one or more states that have a larger overlap with an $S=3/2$, $SU(3)$-octet structure.

\section{Extracting the form factors from ratios of three-point and two-point functions}
\label{sec:ratios}

\begin{figure}

\vspace{-2ex}

\includegraphics[width=0.245\linewidth]{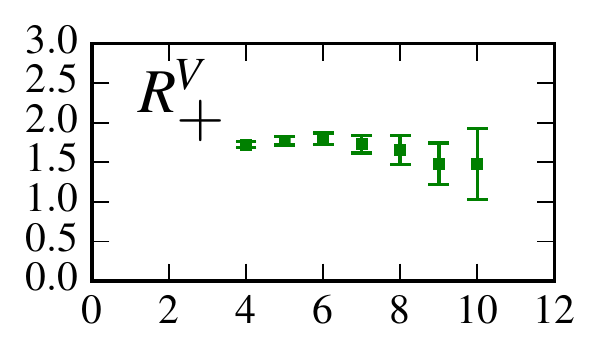} \includegraphics[width=0.245\linewidth]{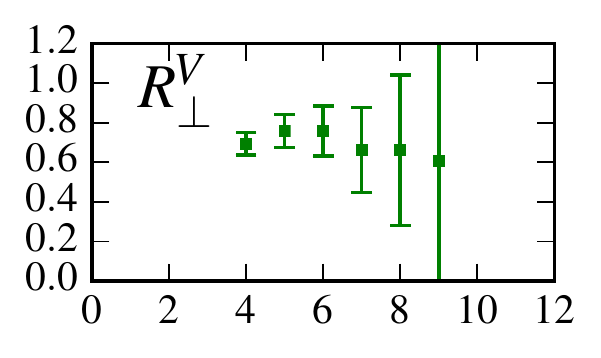} \includegraphics[width=0.245\linewidth]{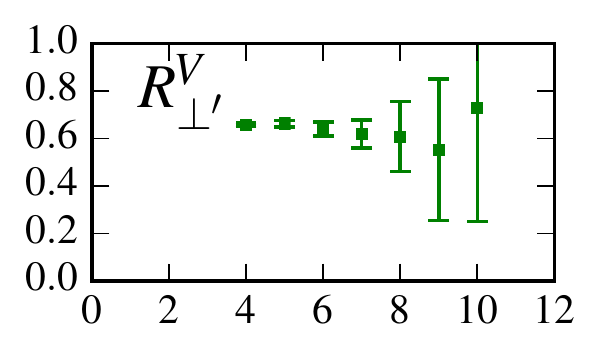} \includegraphics[width=0.245\linewidth]{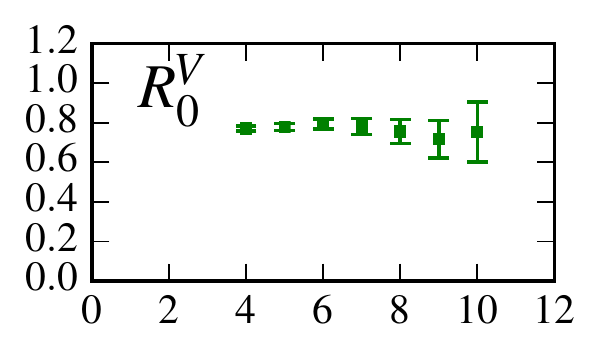}

\includegraphics[width=0.245\linewidth]{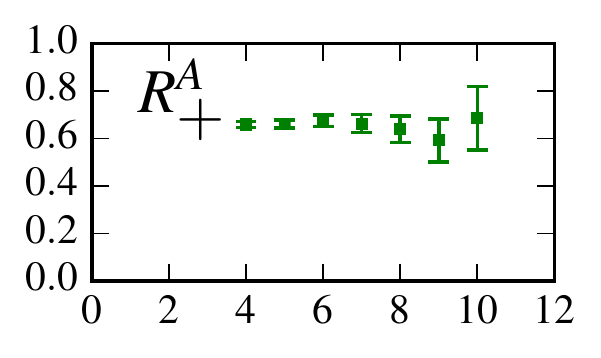} \includegraphics[width=0.245\linewidth]{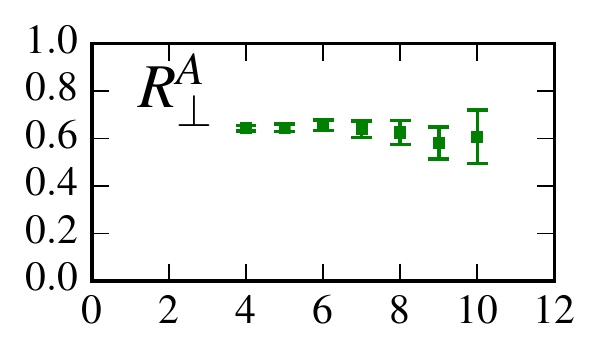} \includegraphics[width=0.245\linewidth]{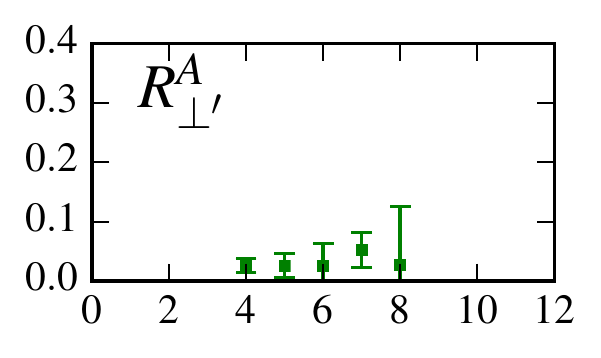} \includegraphics[width=0.245\linewidth]{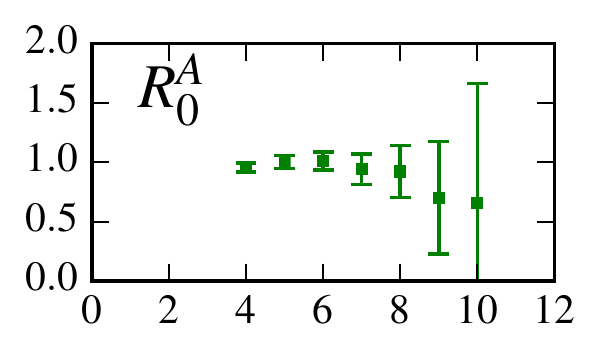}

\vspace{-1ex}

\hspace{0.87\linewidth}{\footnotesize $t/a$}

\vspace{-2.2ex}

\includegraphics[width=0.245\linewidth]{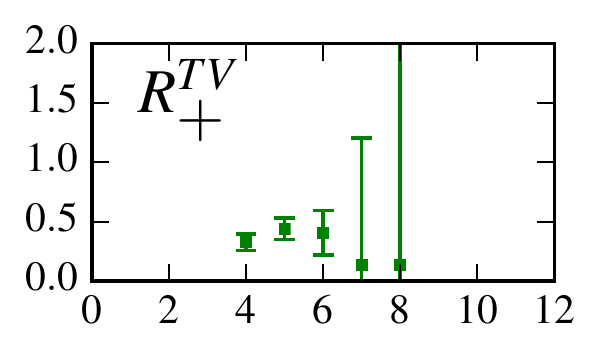} \includegraphics[width=0.245\linewidth]{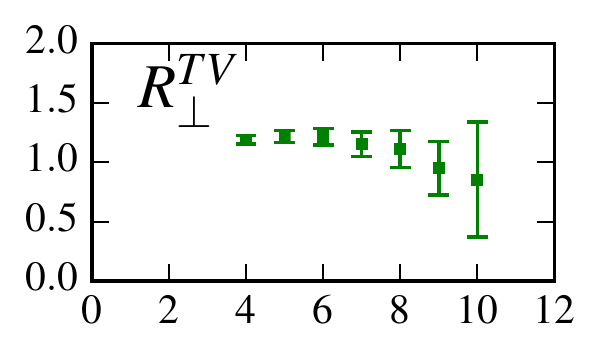} \includegraphics[width=0.245\linewidth]{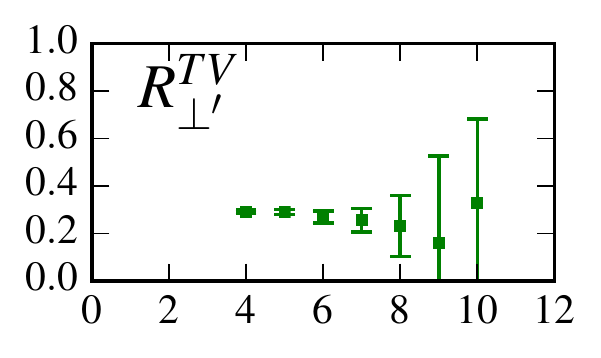}

\includegraphics[width=0.245\linewidth]{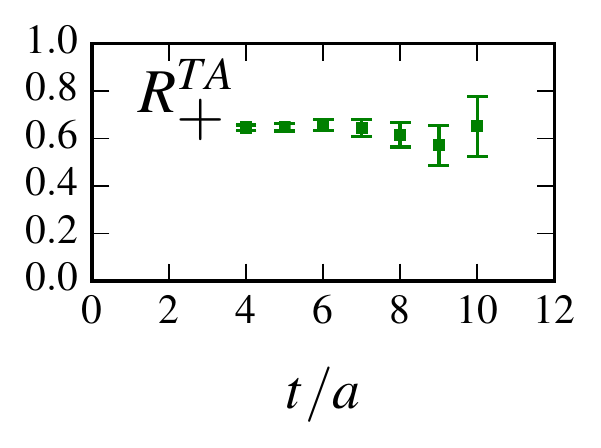} \includegraphics[width=0.245\linewidth]{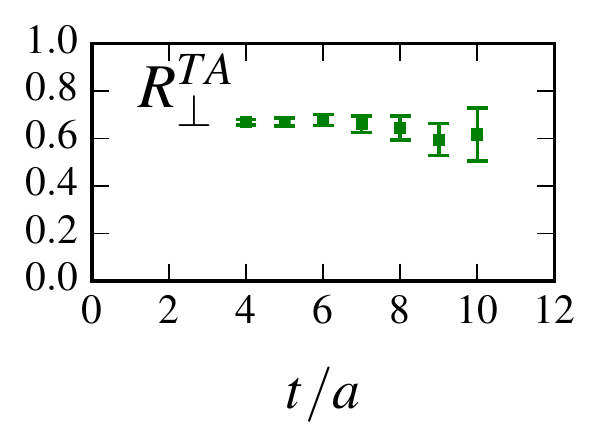} \includegraphics[width=0.245\linewidth]{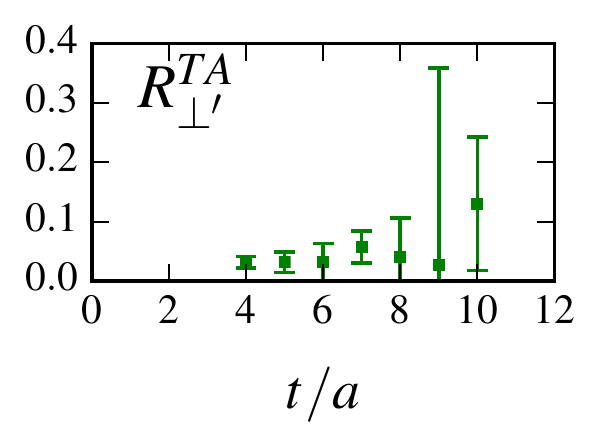}

\vspace{-1ex}

\caption{\label{fig:ratios}Preliminary results for the functions $R_i^X(\mathbf{p}, t)$, defined in Eqs.~(\protect\ref{eq:R0V})-(\protect\ref{eq:RperpprimeV}) for the vector current and similarly for the other currents, at the $\Lambda_b$-momentum $\mathbf{p}=(0,0,3)\frac{2\pi}{L}$. These
 functions become equal to the $\Lambda_b \to \Lambda(1520)$ helicity form factors at the given momentum for large source-sink separation $t$. The results shown here are from 77 configurations (with AMA).}
 \end{figure}

To determine the $\Lambda_b\to\Lambda(1520)$ form factors, we compute three-point functions
\begin{eqnarray}
C^{(3,{\rm fw})}_{j\,\gamma\,\delta}(\mathbf{p},\:\Gamma, t,\:t^\prime) &=& \sum_{\mathbf{y},\mathbf{z}} e^{-i\mathbf{p}\cdot(\mathbf{y}-\mathbf{z})} \left\langle \Lambda_{j\gamma}^{({\rm new})}(x_0,\mathbf{x})\:\: J_\Gamma(x_0-t+t^\prime,\mathbf{y})\:\: \overline{\Lambda}_{b\delta} (x_0-t,\mathbf{z}) \right\rangle,
\end{eqnarray}
where $J_\Gamma=\rho_\Gamma\sqrt{Z_V^{(ss)} Z_V^{(bb)}} \left[ \bar{s}\: \Gamma\: b + a\, d_1\,\bar{s}\: \Gamma\: \bss{\gamma}\cdot\bss{\nabla}  b \right]$ is the renormalized and $\mathcal{O}(a)$-improved $b\to s$ current,
$\Lambda_{b\delta}=\epsilon^{abc}\:(C\gamma_5)_{\alpha\beta}\:\tilde{u}^a_\alpha\:\tilde{d}^b_\beta\: \tilde{b}^c_\delta$ is the interpolating field for the $\Lambda_b$, $\mathbf{p}$ is the momentum of the $\Lambda_b$, and $t$ is
the source-sink separation. The bottom quark is implemented with the relativistic heavy-quark action of Ref.~\cite{Aoki:2012xaa}. Using also the time-reversed backward three-point function and the $\Lambda^*$ and $\Lambda_b$ two-point functions,
we form the ratios
\begin{eqnarray}
 {\mathscr{R}^{jk\mu\nu}(\mathbf{p},t,t^\prime)^X} &=& \frac{ \mathrm{Tr}\Big[ P^{jl} \:  C^{(3,{\rm fw})}_l(\mathbf{p},\Gamma_X^\mu, t, t^\prime) \:\: (\slashed{p}+m_{\Lambda_b}) \:\: C^{(3,{\rm bw})}_{m}(\mathbf{p},\Gamma_X^\nu, t, t-t^\prime) \: P^{mk} \Big] }{\mathrm{Tr}\Big[P^{lm} \: C^{(2,\Lambda^*)}_{lm}(t) \Big]\mathrm{Tr}\Big[(\slashed{p}+m_{\Lambda_b}) \:\: C^{(2,\Lambda_b)}(\mathbf{p},t)  \Big]},
\end{eqnarray}
where $X\in\{V,A,TV,TA\}$ and $\Gamma_V^\mu=\gamma^\mu$, $\Gamma_A^\mu=\gamma^\mu\gamma_5$, $\Gamma_{TV}^\mu=i\sigma^{\mu\nu}q_\nu$, $\Gamma_{TA}^\mu=i\sigma^{\mu\nu}\gamma_5q_\nu$. We then contract with the timelike, longitudinal, and transverse polarization vectors
\begin{equation}
 \epsilon^{(0)} = (\,q^0,\: \mathbf{q}\,), \hspace{3ex}
 \epsilon^{(+)} = (\,|\mathbf{q}|,\: (q^0/|\mathbf{q}|)\mathbf{q}\,), \hspace{3ex}
 \epsilon^{(\perp,\,j)} = (\,0,\: \mathbf{e}_j \times \mathbf{q}\,)
\end{equation}
as follows:
\begin{eqnarray}
{\mathscr{R}_{0}^X(\mathbf{p},t,t^\prime)} &=& g_{jk}\, {\epsilon^{(0)}_\mu  \epsilon^{(0)}_\nu} \, {\mathscr{R}^{jk\mu\nu}(\mathbf{p},t,t^\prime)^X}, \\
{\mathscr{R}_{+}^X(\mathbf{p},t,t^\prime)} &=&  g_{jk}\,{\epsilon^{(+)}_\mu  \epsilon^{(+)}_\nu} \, {\mathscr{R}^{jk\mu\nu}(\mathbf{p},t,t^\prime)^X}, \\
{\mathscr{R}_{\perp}^X(\mathbf{p},t,t^\prime)} &=&  p_j\,p_k\,{\epsilon^{(\perp,l)}_\mu  \epsilon^{(\perp,l)}_\nu} \, {\mathscr{R}^{jk\mu\nu}(\mathbf{p},t,t^\prime)^X}, \\
{\mathscr{R}_{\perp^\prime}^X(\mathbf{p},t,t^\prime)} &=& \left[{\epsilon^{(\perp,m)}_j\epsilon^{(\perp,m)}_k} - \frac12\, p_j\, p_k \right]{\epsilon^{(\perp,l)}_\mu  \epsilon^{(\perp,l)}_\nu} \, {\mathscr{R}^{jk\mu\nu}(\mathbf{p},t,t^\prime)^X}.
\end{eqnarray}
Up to excited-state contamination that is suppressed at large time separations, these quantities are equal to the squares of the individual helicity form factors times known kinematic factors.
For example, in the case of the vector current we obtain the helicity form factors by computing
\begin{eqnarray}
 {R_0^V(\mathbf{p}, t)} &=& \sqrt{\frac{12\, E_{\Lambda_b}m_{\Lambda_b}^2 \:\:{\mathscr{R}_{0}^V(\mathbf{p},t,t/2)}}{(E_{\Lambda_b}-m_{\Lambda_b})(m_{\Lambda_b}-m_\Lambda)^2(E_{\Lambda_b}+m_{\Lambda_b})^2}} \:\:=\:\: {f_0} \:+\: (\text{excited-state contribs.}), \hspace{2ex} \label{eq:R0V} \\
 {R_+^V(\mathbf{p}, t)} &=& \sqrt{\frac{12\, E_{\Lambda_b}m_{\Lambda_b}^2 \:\:{\mathscr{R}_{+}^V(\mathbf{p},t,t/2)}}{(E_{\Lambda_b}+m_{\Lambda_b})(m_{\Lambda_b}+m_\Lambda)^2(E_{\Lambda_b}-m_{\Lambda_b})^2}} \:\:=\:\: {f_+} \:+\: (\text{excited-state contribs.}), \hspace{2ex} \\
 {R_\perp^V(\mathbf{p}, t)} &=& \sqrt{-\frac{9\, E_{\Lambda_b}m_{\Lambda_b}^2  \:\:{\mathscr{R}_{\perp}^V(\mathbf{p},t,t/2) }}{(E_{\Lambda_b}-m_{\Lambda_b})^4(E_{\Lambda_b}+m_{\Lambda_b})^3}}  \:\:=\:\: {f_\perp} \:+\: (\text{excited-state contribs.}),  \\
 {R_{\perp^\prime}^V(\mathbf{p}, t)} &=& \sqrt{-\frac{2\, E_{\Lambda_b}m_{\Lambda_b}^2  \:\:{\mathscr{R}_{\perp^\prime}^V(\mathbf{p},t,t/2)} }{(E_{\Lambda_b}-m_{\Lambda_b})^4(E_{\Lambda_b}+m_{\Lambda_b})^3}}  \:\:=\:\: {f_{\perp^\prime}} \:+\: (\text{excited-state contribs.}). \label{eq:RperpprimeV}
\end{eqnarray}
Preliminary numerical results for these quantities for
all 14 helicity form factors at momentum $\mathbf{p}=(0,0,3)\frac{2\pi}{L}$ are shown in Fig.~\ref{fig:ratios}. Reasonably good signals are obtained for most form factors.

\vspace{-1ex}

\FloatBarrier
\section{Next steps}
\FloatBarrier

\vspace{-1ex}

\begin{figure}
\hfill \includegraphics[width=0.43\linewidth]{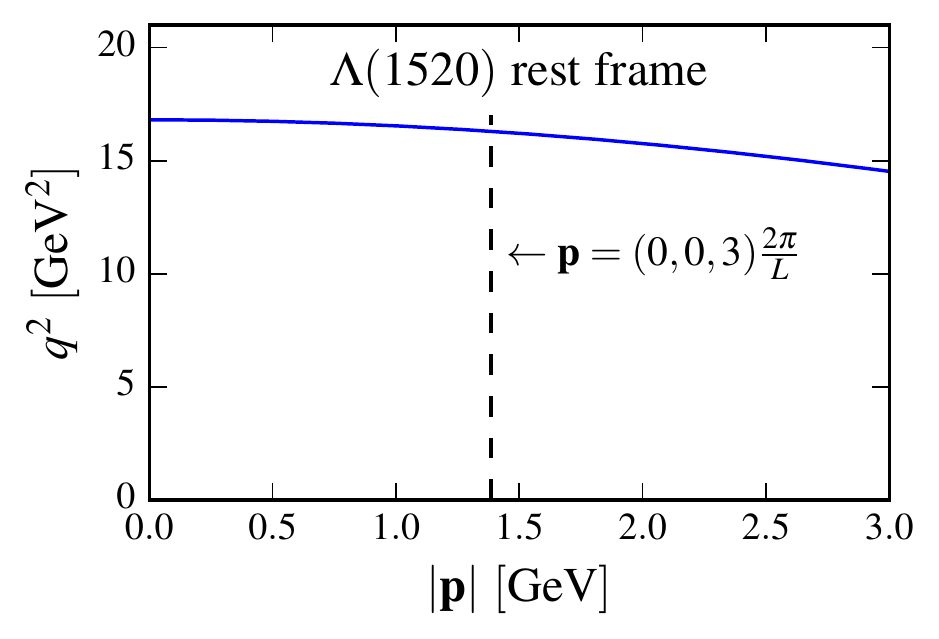}   \hfill \includegraphics[width=0.43\linewidth]{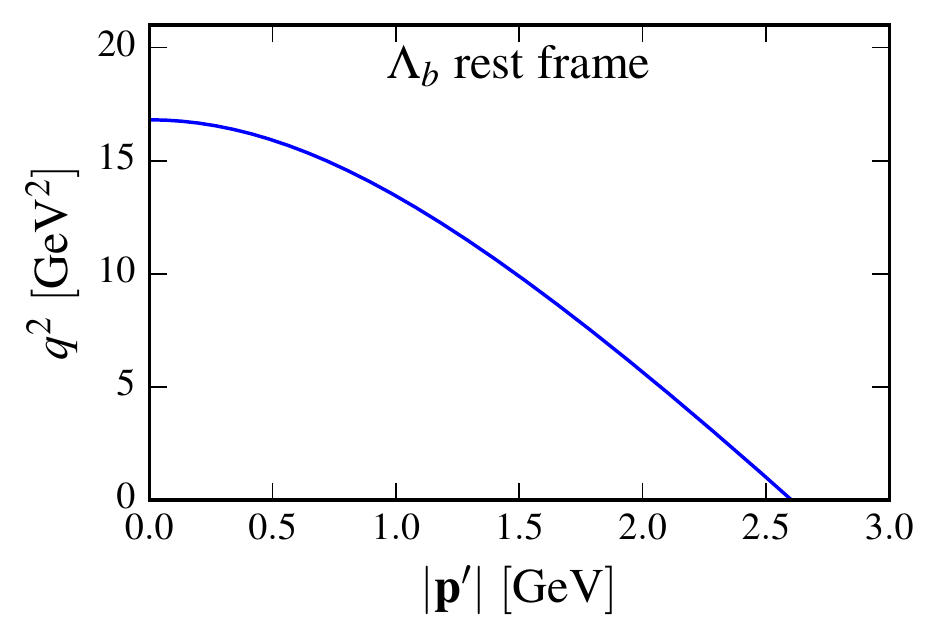} \hfill \null

\vspace{-2ex}

\caption{\label{fig:qsqr}Value of $q^2$ as a function of the $\Lambda_b$ momentum in the $\Lambda(1520)$ rest frame (left), and as a function of the $\Lambda(1520)$ momentum in the $\Lambda_b$ rest frame (right). }
\end{figure}

The drawback of working in the $\Lambda(1520)$ rest frame is that very large $\Lambda_b$ momenta are required to appreciably move $q^2$ away from $q^2_{\rm max}=(m_{\Lambda_b}-m_{\Lambda^*})^2$, as illustrated
in Fig.~\ref{fig:qsqr}. With the relativistic heavy-quark action used so far, this introduces potentially large heavy-quark discretization errors. We therefore plan to perform additional calculations
in which the $b$ quark is implemented with moving NRQCD \cite{Horgan:2009ti}, which will allow us to reach much higher momenta. We also plan to substantially increase statistics and add two
ensembles to study the lattice-spacing and light-quark-mass dependence of the results.

\clearpage

\noindent \textbf{Acknowledgments:} This work is supported by National Science Foundation Grant Number PHY-1520996,
and by the RHIC Physics Fellow Program of the RIKEN BNL Research Center. High-performance computing resources were
provided by XSEDE (supported by National Science Foundation Grant Number OCI-1053575)
and NERSC (supported by U.S.~Department of Energy Grant Number DE-AC02-05CH11231).

\end{document}